\documentclass[%
 aip,
 amsmath,amssymb,
 reprint,%
]{revtex4-1}

\usepackage{graphicx}% Include figure files
\usepackage{dcolumn}% Align table columns on decimal point
\usepackage{bm}% bold math
\usepackage{color}

\usepackage[utf8]{inputenc}
\usepackage[T1]{fontenc}
\usepackage{mathptmx}
\usepackage{etoolbox}

\UseRawInputEncoding

\makeatletter
\def\@email#1#2{%
 \endgroup
 \patchcmd{\titleblock@produce}
  {\frontmatter@RRAPformat}
  {\frontmatter@RRAPformat{\produce@RRAP{*#1\href{mailto:#2}{#2}}}\frontmatter@RRAPformat}
  {}{}
}%
\makeatother

\begin{document}

\preprint{AIP/123-QED}

\title{Suppression of atomic displacive excitation in photo-induced A$_{\mathrm{1g}}$ phonon mode of bismuth unveiled by low-temperature time-resolved x-ray diffraction}
% Force line breaks with \\

\author{Yuya Kubota}
\email{kubota@spring8.or.jp}
\affiliation{RIKEN SPring-8 Center, 1-1-1 Kouto, Sayo, Hyogo 679-5148, Japan}%Lines break automatically or can be forced with \\
 
\author{Yoshikazu Tanaka}
\affiliation{RIKEN SPring-8 Center, 1-1-1 Kouto, Sayo, Hyogo 679-5148, Japan}

\author{Tadashi Togashi}
\affiliation{Japan Synchrotron Radiation Research Institute (JASRI), 1-1-1 Kouto, Sayo, Hyogo 679-5198, Japan}
\affiliation{RIKEN SPring-8 Center, 1-1-1 Kouto, Sayo, Hyogo 679-5148, Japan}

\author{Tomio Ebisu}
\affiliation{RIKEN SPring-8 Center, 1-1-1 Kouto, Sayo, Hyogo 679-5148, Japan}

\author{Kenji Tamasaku}
\affiliation{RIKEN SPring-8 Center, 1-1-1 Kouto, Sayo, Hyogo 679-5148, Japan}

\author{Hitoshi Osawa}
\affiliation{Japan Synchrotron Radiation Research Institute (JASRI), 1-1-1 Kouto, Sayo, Hyogo 679-5198, Japan}

\author{Tetsuya Wada}
\affiliation{Institute for Solid State Physics, The University of Tokyo, Kashiwa, Chiba 277-8581, Japan}

\author{Osamu Sugino}
\affiliation{Institute for Solid State Physics, The University of Tokyo, Kashiwa, Chiba 277-8581, Japan}

\author{Iwao Matsuda}
\affiliation{Institute for Solid State Physics, The University of Tokyo, Kashiwa, Chiba 277-8581, Japan}

\author{Makina Yabashi}
\affiliation{RIKEN SPring-8 Center, 1-1-1 Kouto, Sayo, Hyogo 679-5148, Japan}
\affiliation{Japan Synchrotron Radiation Research Institute (JASRI), 1-1-1 Kouto, Sayo, Hyogo 679-5198, Japan}

\date{\today}% It is always \today, today,

\begin{abstract}
An ultrafast atomic motion of a photo-induced coherent phonon of bismuth at low temperatures was directly observed with time-resolved x-ray diffraction.
A cryostat with a window that is transparent to both optical lasers and x-rays enabled versatile diffraction measurements in a wide temperature range including below  $10$~K.
It is found that an atomic displacement in a fully symmetric A$_{\mathrm{1g}}$ phonon mode is suppressed at low temperatures.
This result indicates the displacive excitation process is suppressed in the phonon generation with decreasing temperature.
\end{abstract}

\maketitle

%\begin{quotation}
%The ``lead paragraph'' is encapsulated with the \LaTeX\ 
%\verb+quotation+ environment and is formatted as a single paragraph before the first section heading. 
%(The \verb+quotation+ environment reverts to its usual meaning after the first sectioning command.) 
%Note that numbered references are allowed in the lead paragraph.
%
%The lead paragraph will only be found in an article being prepared for the journal \textit{Chaos}.
%\end{quotation}
%\section{\label{sec:level1}First-level heading:\protect\\ The line break was forced \lowercase{via} \textbackslash\textbackslash}

Controlling quantum phenomena in materials has been one of the central challenges in condensed matter physics.
A well-promising technique is stimulating materials by external fields to excite into non-equilibrium states, where various exotic physical properties emerge.
Optical pulses have been widely used as excitation sources to induce non-equilibrium quantum phenomena, such as ultrafast phase transitions, magnetic order transitions, and photo-induced superconductivities~\cite{Miyano1997, Kirilyuk2010, Buzzi2018}.
Since an ultrashort pulse triggers physical events in materials before reaching the thermal equilibrium, it has opened various non-equilibrium dynamics in the quantum phases that exist only at temperatures much lower than room temperature.

Time-resolved x-ray diffraction (XRD) combined with optical pump lasers is a powerful method to investigate non-equilibrium quantum phenomena.
Furthermore, x-ray free-electron lasers (XFELs)~\cite{Kondratenko1980, Bonifacio1984} enable the unveiling of their ultrafast dynamics in femtosecond timescale~\cite{Buzzi2018} such as atomic motions of photo-induced coherent phonon~\cite{Nelson1982, Cheng1990, Merlin1997}.
However, performing pump-probe XRD experiments has been mostly limited to a temperature range above $\sim 100$~K under ambient sample environments.
This is because x-ray windows of standard cryostats used for low-temperature XRD, such as beryllium, carbon, and aluminum are opaque to optical lasers.
Although cryostats in vacuum chambers directly connected to x-ray beamlines can achieve low temperatures below $100$~K, experimental conditions are severely restricted by limitations of sample motion and $2 \theta$ angular range.
A new cooling system with a suitable window material is highly required to investigate quantum phenomena under various conditions.

In this Letter, we report a notable development of the system with high versatility for pump-probe XRD at a cryogenic temperature, and direct observation of the atomic motion of bismuth (Bi) in the photo-induced coherent phonon at $T = 9$~K.
Since Bi has a strong electron-lattice coupling, it has been a model material for investigating the photo-induced coherent phonon over a wide temperature range with an optical pump-optical probe (OPOP) method~\cite{Cheng1990, Hase1998, Hase2002, Misochko2004, Misochko2006, Ishioka2006, Li2013}.
However, previous studies have reported the atomic motion indirectly through data interpretation based on the electron-lattice interaction.
Due to a lack of direct data, the issue of what process is responsible for phonon generation has remained controversial.
To examine this question, it is essentially important to directly observe the lattice dynamics on a femtosecond time scale with pump-probe XRD.
Our results show a suppression of the atomic displacement in the photo-induced coherent phonon at low temperatures, indicating the contribution of the displacive excitation process decreases in the phonon generation.

\begin{figure}
\begin{center}
\includegraphics[width=8cm]{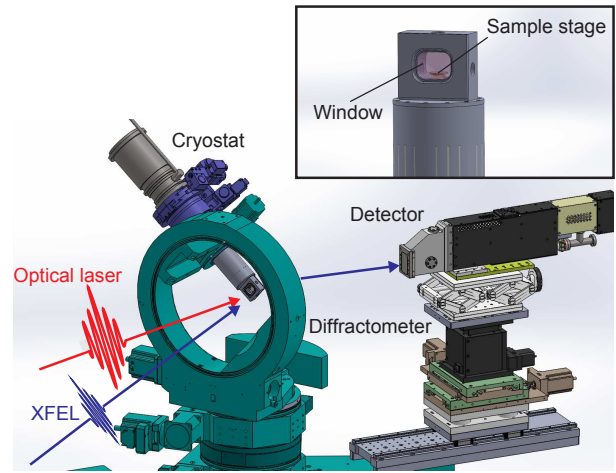}% Here is how to import EPS art
\caption{Schematic of the pump-probe XRD experiment at low temperatures.
A cryostat with a transparent polyimide film window is mounted on a standard four-circle diffractometer.
The diffracted x-ray beam is detected with the MPCCD on the $2 \theta$ arm.
The inset shows a magnified drawing of the cryostat around the sample position.
A transparent polyimide film is used as a window material on the incoming side, while an aluminum-evaporated normal polyimide film, which transmits only x-rays, is used on the outgoing side.}
\label{fig1}
\end{center}
\end{figure}
\begin{figure}
\begin{center}
\includegraphics[width=8cm]{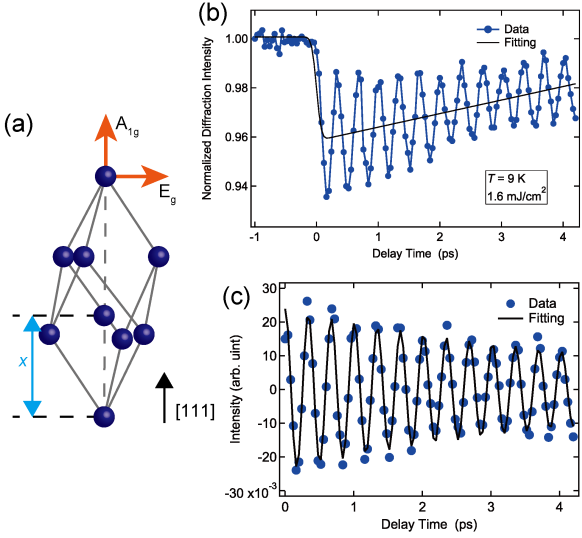}% Here is how to import EPS art
\caption{
(a) Structure of Bi unit cell.
The blue arrow represents the length of two atomic positions, $x$, in a unit of the diagonal length along the $[1 1 1]$ axis.
The atoms move parallel and perpendicular to the $[1 1 1]$ axis in the A$_{\mathrm{1g}}$ and E$_{\mathrm{g}}$ phonon modes, respectively, as shown by the orange arrows.
(b) A typical result of the Bi $1 1 1$ diffraction intensity variation as a function of the delay time obtained at $T = 9$~K and fluence of $1.6$~mJ/cm$^2$, shown as the blue circles.
The blue line is a guide to the eye.
The black solid line represents a fitting curve of the non-oscillatory component with an exponential decay function convoluted with a Gaussian function.
(c) Oscillatory component obtained by subtracting the non-oscillatory component from (b), shown as the blue circles.
The black solid line represents a fitting exponentially decaying cosine curve.
}
\label{fig2} 
\end{center}
\end{figure}

A pump-probe XRD experiment was performed at BL3 of SACLA~\cite{Ishikawa2012}.
Figure~\ref{fig1}(a) shows a schematic drawing of the experimental setup.
To achieve a cryogenic temperature below $100$~K, we employed a highly transparent polyimide film for infrared and visible light with a thickness of $50$~$\mu$m ($> 90$\% at wavelengths above $400$~nm) as a window material.
It enables us a pump-probe XRD experiment at low temperatures below $10$~K.
The cryostat was mounted on a standard four-circle diffractometer to collect XRD signals over a wide angular range, e.g. up to $\sim 100$~degrees of $2 \theta$.
A $53$-nm-thick Bi film epitaxially grown on a Si~$(1 1 1)$ substrate~\cite{Kammler2005} was excited with 800-nm optical laser pulses with a full-width-at-half-maximum (FWHM) duration of $\sim 40$~fs~\cite{Togashi2020}.
The photon energy $h\rm \nu$ and pulse duration of the XFEL probe beam were $10$~keV with a bandwidth of $\sim 1$~eV and $\sim 7$~fs (FWHM), respectively~\cite{Inubushi2012, Inubushi2017, Inoue2018}.
The XFEL beam was focused to $100$~$\mu$m in diameter (FWHM) with compound refractive lenses, while the optical laser was focused to $870$~$\mu$m in diameter (FWHM).
The angle between the optical laser and the XFEL beams was $1.8$ degrees.
The XFEL intensity of Bi $1 1 1$ diffraction in a photo-excited state was detected with a multi-port charge-coupled device (MPCCD)~\cite{Kameshima2014} as a function of the delay time of the XFEL pulse from the optical laser pulse.
The intensity variation of Bi $1 1 1$ diffraction probes the A$_{\mathrm{1g}}$ phonon mode oscillation along the $[1 1 1]$ direction as shown in Fig.~\ref{fig2}(a).

Figure~\ref{fig2}(b) shows a typical result of the Bi $1 1 1$ diffraction intensity variation as a function of the delay time ($t$) associated with the A$_{\mathrm{1g}}$ coherent phonon mode.
This result was obtained at $T = 9$~K and an absorbed excitation fluence of $1.6$~mJ/cm$^2$, which value was evaluated taking into account the reflectivity of $33$\% in the case of Bi at an incident angle of $80$ degrees for $p$ polarization in this study~\cite{Johnson2013}.
We confirmed that the lattice constant was unchanged in a time range up to at least $4$~ps after a photo-excitation, judging from the position of the XRD signal observed on the MPCCD.
Note that the variations of diffraction intensity represent average values over the sample thickness because the XFEL beam is almost uniformly transmitted by the 53-nm thickness of Bi.

To obtain the information on the A$_{\mathrm{1g}}$ coherent phonon mode, we firstly fit the non-oscillatory component with an exponential decay function convoluted with a Gaussian function as shown by the black solid line in Fig.~\ref{fig2}(b).
We secondly obtain the oscillatory component after the subtraction of the non-oscillatory component from the total intensity variation as shown in Fig.~\ref{fig2}(c).
The frequency of photo-induced A$_{\mathrm{1g}}$ coherent phonon mode is $2.95$~THz at $T = 9$~K and the fluence of $1.6$~mJ/cm$^2$ obtained from fitting with an exponentially decaying cosine curve.
This result is almost consistent with the previous OPOP study~\cite{Misochko2004}.
The value of frequency is the average phonon frequency over the time range of $0$ to $\sim 4$~ps because the chirp effect was negligible as shown in Fig.~\ref{fig2}(c).

\begin{figure}
\begin{center}
\includegraphics[width=8cm]{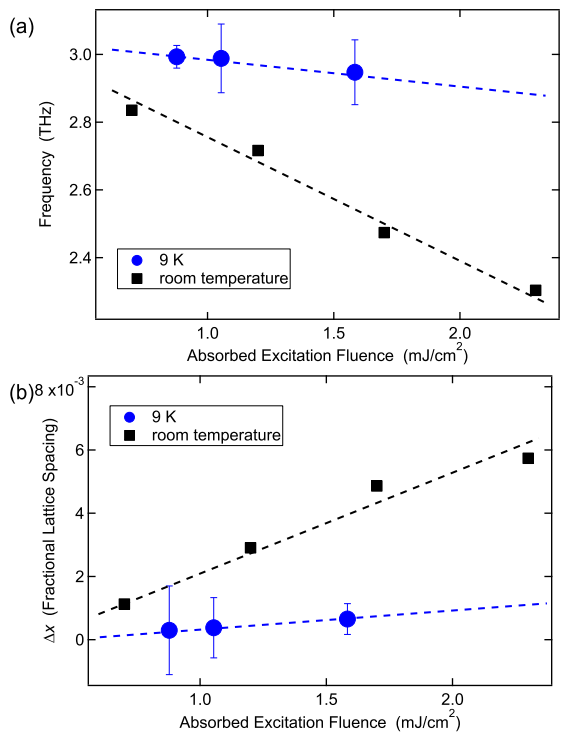}% Here is how to import EPS art
\caption{
(a) Frequency of the A$_{\mathrm{1g}}$ phonon mode of Bi obtained by fitting and (b) the atomic displacement $\Delta x$ obtained from the variation of diffraction intensity around $t = 0$ at $T =9$~K as a function of absorbed excitation fluence (blue circles), compared with those at room temperature reported by Ref.~\cite{Fritz2007} (black squares).
Error bars are 95\% confidence intervals.
The dashed lines are guides to the eye.
}
\label{fig3} 
\end{center}
\end{figure}

We measured the fluence dependence at $T = 9$~K to investigate the difference in behavior between the low and room temperatures.
Figure~\ref{fig3}(a) shows the A$_{\mathrm{1g}}$ phonon frequency as a function of fluence, which is compared with the published data measured at room temperature~\cite{Fritz2007}.
Although we observed a decrease in the phonon frequency as the fluence increases at $T = 9$~K as well as at room temperature~\cite{Fritz2007}, the reduction at $T = 9$~K is much smaller.
This behavior is consistent with the previous result obtained in the OPOP experiment~\cite{Misochko2004}.

We further investigate the atomic motion, which is inaccessible to the OPOP method, by analyzing the XRD data.
The space group of Bi is $\#166$ (R$\bar{3}$m) and its structure factor for the $h k l$ Bragg diffraction, $F_{h k l}$, has the form
\begin{eqnarray}
F_{h k l} &=& f_{\mathrm{Bi}} \left\{ 1 + \exp \left[ 2 \pi i \left( h + k + l \right) x \right] \right\} \nonumber \\
&=& 2 f_{\mathrm{Bi}} \exp \left[ \pi i \left( h + k + l \right) x \right] \cos \left[ \pi \left( h + k + l \right) x \right],
\label{F}	
\end{eqnarray}
where $f_{\mathrm{Bi}}$ is the atomic scattering factor for Bi, and $x$ is the atomic position in the trigonal unit cell along the $[1 1 1]$ direction in a unit of the hexagonal unit-cell length, $c$, as shown in Fig.~\ref{fig2}(a).
The variation of the normalized $1 1 1$ diffraction intensity, $I (t)/I (0)$, caused by the atomic displacement along $[1 1 1]$ direction, is represented by
\begin{equation}
\frac{I \left( t \right)}{I \left( 0 \right)} = \frac{\left| F_{1 1 1} \left( t \right) \right| ^2}{\left| F_{1 1 1} \left( 0 \right) \right| ^2} = \frac{\cos^2 \left[ 3 \pi x \left( t \right) \right]}{\cos^2 \left[ 3 \pi x \left( 0 \right) \right]}.
\label{I}
\end{equation}
Figure~\ref{fig3}(b) shows fluence dependence of the magnitude of the deviation of $x$ from the equilibrium point, i.e. $\Delta x = x - x \left( 0 \right)$, at $T = 9$~K.
The values are obtained from the minima of the fitting curves to the non-oscillatory component around $t = 0$ shown in Fig.~\ref{fig2}(b).
For the low-temperature condition, the equilibrium values of $x \left( 0 \right)$ and $c$ before a photo-excitation were $0.46814$ and $11.797$~\AA, respectively~\cite{Cucka1962}.
The crystal structure of Bi is a result of Peierls distortion from a high-symmetry cubic structure.
This distortion deviates $x$ from $0.5$ of the cubic structure.
The decrease of the diffraction intensity shown in Fig.~\ref{fig2}(b) indicates that quasi-equilibrium atomic position moves toward $x = 0.5$ because the $1 1 1$ diffraction is forbidden for $x = 0.5$~\cite{Fritz2007}.
As the fluence increases, $x$ gradually increases toward $x = 0.5$, which implies the Peierls distortion decreases.
However, the value of $\Delta x$ observed at $T = 9$~K is significantly smaller than that observed at room temperature~\cite{Fritz2007}, as shown in Fig.~\ref{fig3}(b).
This result is consistent with the small change in the phonon frequency as the fluence increases at $T = 9$~K [Fig.~\ref{fig3}(a)]~\cite{Fritz2007}.
Note that we also measured the coherent phonon at room temperature with the fluence of $1.2$~mJ/cm$^2$ and confirmed that the phonon frequency and $\Delta x$ match with the published data~\cite{Fritz2007}, as shown in the supplementary material.

\begin{figure}
\begin{center}
\includegraphics[width=8cm]{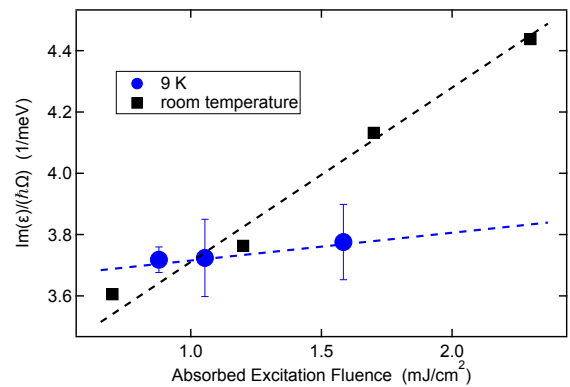}% Here is how to import EPS art
\caption{
$\mathrm{Im} ( \varepsilon ) / ( \hbar \Omega )$ of the A$_{\mathrm{1g}}$ phonon mode of Bi as a function of absorbed excitation fluence at $T = 9$~K (blue circles) and room temperature (black squares) using the values in Fig.~\ref{fig3}(a).
The dashed lines are guides to the eye.
}
\label{fig4}
\end{center}
\end{figure}
The time-resolved XRD observation of the small deviation $\Delta x$ at $T = 9$~K provides a notable insight into the phonon-generation mechanism that has been controversial.
First, we note that the deviation corresponds to the fully symmetric A$_{\mathrm{1g}}$ phonon mode, which has been observed predominantly at room temperature.
The generation of this mode was explained by the displacive excitation of coherent phonons (DECP) model~\cite{Zeiger1992}.
In the DECP model, photo-excited carriers induce a shift of the equilibrium atomic position followed by atomic oscillations at the new position.
In the meanwhile, there is another mode of photo-induced coherent phonon, the doubly degenerate E$_{\mathrm{g}}$ phonon mode, where atoms move in the directions perpendicular to $[1 1 1]$ [Fig.~\ref{fig2}(a)]~\cite{Misochko2006, Ishioka2006, Li2013, Johnson2013}.
The E$_{\mathrm{g}}$ phonon generation was explained by the impulsive stimulated Raman scattering (ISRS) model, where an ultrafast optical pulse having a broad energy spectrum impulsively induces stimulated Raman scattering resulting in atomic oscillations at the equilibrium position~\cite{Yan1985}.
In the literature, the E$_{\mathrm{g}}$ phonon mode was reported to be generated at low temperatures~\cite{Misochko2006, Ishioka2006, Li2013, Johnson2013}.
This is ascribed to impulsive excitation, which is considered to play a more influential role as the temperature is decreased.
Furthermore, the previous OPOP and spontaneous Raman scattering measurements~\cite{Li2013} concluded that the DECP and ISRS mechanisms contribute comparably even in the A$_{\mathrm{1g}}$ phonon mode at low temperatures.

In this context, the small value of $\Delta x$ indicates that displacive excitation in the A$_{\mathrm{1g}}$ phonon mode is suppressed, while the contribution of the impulsive excitation becomes notable in the coherent phonon generation of Bi with decreasing temperature.
Since the DECP scheme can be interpreted as the resonant case of ISRS, this transfer can be systematically theorized in terms of the Raman scattering process, as developed by Stevens {\it et al.}~\cite{Steves2002}.
The stimulated Raman scattering is described therein by two tensors, consisting of a common real component but different imaginary components.
If the imaginary (real) part dominates, the generation mechanism of the photo-induced coherent phonon is displacive (impulsive).
Therefore, one can interpret our measurement in terms of the temperature-dependent imaginary component when one assumes that the essence of photo-induced coherent phonon generation in Bi is the stimulated Raman scattering process.
The imaginary component, which contributes to the displacive driving force, is proportional to $\mathrm{Im} ( \varepsilon ) / \Omega$, where $\varepsilon$ is the dielectric function, and $\Omega$ is the phonon frequency~\cite{Steves2002}.
Figure~\ref{fig4} shows fluence dependence of $\mathrm{Im} ( \varepsilon ) / \Omega$ at $T = 9$~K and room temperature using the values of $\Omega$ shown in Fig.~\ref{fig3}(a).
The experimental values of $\mathrm{Im} ( \varepsilon )$ reported by Hunderi~\cite{Hunderi1975} were used.
Although the values of $\mathrm{Im} ( \varepsilon ) / \Omega$ at $T = 9$~K and room temperature are comparable in the small fluence region, the values at $T = 9$~K are smaller than those at room temperature in the high fluence region ($> 1$~mJ/cm$^2$).
This is attributable to the fact that $\Omega$ at low temperatures is larger and its smaller change as the fluence increases than those at room temperature, as shown in Fig.~\ref{fig3}(a) and the previous OPOP studies~\cite{Hase1998, Misochko2004, Ishioka2006, Li2013}.
This result indicates that high phonon frequency is equivalent to the small driving force of DECP at low temperatures, which corroborates our result of the little variation of $\Delta x$.
The temperature dependence of $\mathrm{Im} ( \varepsilon ) / \Omega$ might be smaller than that of $\Delta x$.
However, one can expect that redistribution of excitation energy from the A$_{\mathrm{1g}}$ to the E$_{\mathrm{g}}$ mode also contributes to the suppression of $\Delta x$ from the previous report indicating that the electron-phonon coupling of the E$_{\mathrm{g}}$ phonon mode becomes stronger at low temperatures than that at room temperature~\cite{Li2013}.

In conclusion, we developed a system with high versatility for pump-probe XRD at a cryogenic temperature.
A cryostat with a transparent polyimide film window was mounted on a standard four-circle diffractometer, resulting in a wide acceptable range of diffraction angles even at low temperatures.
We directly observed the suppression of atomic displacement in the photo-induced A$_{\mathrm{1g}}$ coherent phonon mode of Bi at $T = 9$~K.
The result indicates that the contribution of the displacive excitation process decreases at low temperatures.
It can be corroborated by the temperature dependence of the imaginary component of the tensor, which describes stimulated Raman scattering.
Our experimental discovery and the phenomenological model become the research targets for the future development of theoretical studies to obtain a full understanding of the generation mechanism of the coherent phonon.
This understanding will lead to the control of phonons using light, which promotes applications such as a conversion of phonon oscillations into coherent electromagnetic energy~\cite{Dekorsy1995} and manipulation of atomic position and phonon to realize high-$T_{\mathrm{c}}$ superconductivities~\cite{Okazaki2018, Gerber2017, Suzuki2019, Mankowsky2014}.
Our cryo-diffraction system with XFEL opens up rich avenues for unexpected discoveries and application developments in condensed matter physics.

This experiment was performed at BL3 of SACLA with the
approval of the Japan Synchrotron Radiation Research Institute (JASRI) (Proposal Nos.~2018B8011, 2019A8065, 2019B8045, and 2020A8070).
The synchrotron radiation experiments were performed to develop and evaluate the system at BL19LXU in SPring-8 with the approval of RIKEN (Proposal Nos.~20180094 and 20190042).
We thank Dr. Suguru Ito for the sample preparation, and Dr. Jun Haruyama and Dr. Takeshi Suzuki for the valuable discussion.

SUPPLEMENTARY MATERIAL

See supplementary material for the experimental results obtained at room temperature.

AUTHOR DECLARATIONS

Conflict of Interest

The authors have no conflicts to disclose.

DATA AVAILABILITY

The data that support the findings of this study are available from the corresponding author upon reasonable request.

%\bibliography{aipsamp}% Produces the bibliography via BibTeX.

\begin{thebibliography}{99}
{

\bibitem{Miyano1997} K. Miyano, T. Tanaka, Y. Tomioka, and Y. Tokura, Phys. Rev. Lett. {\bf 78}, 4257 (1997).

\bibitem{Kirilyuk2010} A. Kirilyuk, A. V. Kimel, and T. Rasing, Rev. Mod. Phys. {\bf 82}, 2731 (2010).

\bibitem{Buzzi2018} M. Buzzi, M. F$\ddot{\mathrm{o}}$rst, R. Mankowsky, and A. Cavalleri, Nat. Rev. Mater. {\bf 3}, 1 (2018).

\bibitem{Kondratenko1980} A. M. Kondratenko and E. L. Saldin, Part. Accel. {\bf 10}, 207 (1980).

\bibitem{Bonifacio1984} R. Bonifacio, C. Pellegrini, and L. M. Narducci, Opt. Commun. {\bf 50}, 373 (1984).

\bibitem{Nelson1982} K. A. Nelson, R. J. D. Miller, D. R. Lutz, and M. D. Fayer, J. Appl. Phys. {\bf 53}, 1144 (1982).

\bibitem{Cheng1990} T. K. Cheng, S. D. Brorson, A. S. Kazeroonian, J. S. Moodera, G. Dresselhaus, M. S. Dresselhaus, and E. P. Ippen, Appl. Phys. Lett. {\bf 57}, 1004 (1990).

\bibitem{Merlin1997} R. Merlin, Solid State Commun. {\bf 102}, 207 (1997).

\bibitem{Hase1998} M. Hase, K. Mizoguchi, H. Harima, S. Nakashima, and K. Sakai, Phys. Rev. B {\bf 58}, 5448 (1998).

\bibitem{Hase2002} M. Hase, M. Kitajima, S. Nakashima, and K. Mizoguchi, Phys. Rev. Lett. {\bf 88}, 067401 (2002).

\bibitem{Misochko2004} O. V. Misochko, M. Hase, K. Ishioka, and M. Kitajima, Phys. Rev. Lett. {\bf 92}, 197401 (2004).

\bibitem{Misochko2006} O. V Misochko, K. Ishioka, M. Hase, and M. Kitajima, J. Phys. Condens. Matter {\bf 18}, 10571 (2006).

\bibitem{Ishioka2006} K. Ishioka, M. Kitajima, and O. V. Misochko, J. Appl. Phys. {\bf 100}, 093501 (2006).

\bibitem{Li2013} J. J. Li, J. Chen, D. A. Reis, S. Fahy, and R. Merlin, Phys. Rev. Lett. {\bf 110}, 047401 (2013).

\bibitem{Ishikawa2012} T. Ishikawa, H. Aoyagi, T. Asaka, Y. Asano, N. Azumi, T. Bizen, H. Ego, K. Fukami, T. Fukui, Y. Furukawa, S. Goto, H. Hanaki, T. Hara, T. Hasegawa, T. Hatsui, A. Higashiya, T. Hirono, N. Hosoda, M. Ishii, T. Inagaki, Y. Inubushi, T. Itoga, Y. Joti, M. Kago, T. Kameshima, H. Kimura, Y. Kirihara, A. Kiyomichi, T. Kobayashi, C. Kondo, T. Kudo, H. Maesaka, X. M. Mar$\acute{\mathrm{e}}$chal, T. Masuda, S. Matsubara, T. Matsumoto, T. Matsushita, S. Matsui, M. Nagasono, N. Nariyama, H. Ohashi, T. Ohata, T. Ohshima, S. Ono, Y. Otake, C. Saji, T. Sakurai, T. Sato, K. Sawada, T. Seike, K. Shirasawa, T. Sugimoto, S. Suzuki, S. Takahashi, H. Takebe, K. Takeshita, K. Tamasaku, H. Tanaka, R. Tanaka, T. Tanaka, T. Togashi, K. Togawa, A. Tokuhisa, H. Tomizawa, K. Tono, S. Wu, M. Yabashi, M. Yamaga, A. Yamashita, K. Yanagida, C. Zhang, T. Shintake, H. Kitamura, and N. Kumagai, Nat. Photonics {\bf 6}, 540 (2012).

\bibitem{Kammler2005} M. Kammler and M. Horn-Von Hoegen, Surf. Sci. {\bf 76}, 56 (2005).

\bibitem{Togashi2020} T. Togashi, S. Owada, Y. Kubota, K. Sueda, T. Katayama, H. Tomizawa, T. Yabuuchi, K. Tono, and M. Yabashi, Appl. Sci. {\bf 10}, 7934 (2020).

\bibitem{Inubushi2012} Y. Inubushi, K. Tono, T. Togashi, T. Sato, T. Hatsui, T. Kameshima, K. Togawa, T. Hara, T. Tanaka, H. Tanaka, T. Ishikawa, and M. Yabashi, Phys. Rev. Lett. {\bf 109}, 144801 (2012).

\bibitem{Inubushi2017} Y. Inubushi, I. Inoue, J. Kim, A. Nishihara, S. Matsuyama, H. Yumoto, T. Koyama, K. Tono, H. Ohashi, K. Yamauchi, and M. Yabashi, Appl. Sci. {\bf 7}, 584 (2017).

\bibitem{Inoue2018} I. Inoue, T. Hara, Y. Inubushi, K. Tono, T. Inagaki, T. Katayama, Y. Amemiya, H. Tanaka, and M. Yabashi, Phys. Rev. Accel. Beams {\bf 21}, 080704 (2018).

\bibitem{Kameshima2014} T. Kameshima, S. Ono, T. Kudo, K. Ozaki, Y. Kirihara, K. Kobayashi, Y. Inubushi, M. Yabashi, T. Horigome, A. Holland, K. Holland, D. Burt, H. Murao, and T. Hatsui, Rev. Sci. Instrum. {\bf 85}, 033110 (2014).

\bibitem{Johnson2013} S. L. Johnson, P. Beaud, E. M$\ddot{\mathrm{o}}$hr-Vorobeva, A. Caviezel, G. Ingold, and C. J. Milne, Phys. Rev. B {\bf 87}, 054301 (2013).

\bibitem{Fritz2007} D. M. Fritz, D. A. Reis, B. Adams, R. A. Akre, J. Arthur, C. Blome, P. H. Bucksbaum, A. L. Cavalieri, S. Engemann, S. Fahy, R. W. Falcone, P. H. Fuoss, K. J. Gaffney, M. J. George, J. Hajdu, M. P. Hertlein, P. B. Hillyard, M. Horn-von Hoegen, M. Kammler, J. Kaspar, R. Kienberger, P. Krejcik, S. H. Lee, A. M. Lindenberg, B. McFarland, D. Meyer, T. Montagne, É. D. Murray, A. J. Nelson, M. Nicoul, R. Pahl, J. Rudati, H. Schlarb, D. P. Siddons, K. Sokolowski-Tinten, T. Tschentscher, D. von der Linde, and J. B. Hastings, Science {\bf 315}, 633 (2007).

\bibitem{Cucka1962} P. Cucka and C. S. Barrett, Acta Crystallogr. {\bf 15}, 865 (1962).

\bibitem{Zeiger1992} H. J. Zeiger, J. Vidal, T. K. Cheng, E. P. Ippen, G. Dresselhaus, and M. S. Dresselhaus, Phys. Rev. B {\bf 45}, 768 (1992).

\bibitem{Yan1985} Y. X. Yan, E. B. Gamble, and K. A. Nelson, J. Chem. Phys. {\bf 83}, 5391 (1985).

\bibitem{Steves2002} T. E. Stevens, J. Kuhl, and R. Merlin, Phys. Rev. B {\bf 65}, 144304 (2002).

\bibitem{Hunderi1975} O. Hunderi, J. Phys. F Met. Phys. {\bf 5}, 2214 (1975).

\bibitem{Dekorsy1995} T. Dekorsy, H. Auer, C. Waschke, H. J. Bakker, H. G. Roskos, H. Kurz, V. Wagner, and P. Grosse, Phys. Rev. Lett. {\bf 74}, 738 (1995).

\bibitem{Okazaki2018} K. Okazaki, H. Suzuki, T. Suzuki, T. Yamamoto, T. Someya, Y. Ogawa, M. Okada, M. Fujisawa, T. Kanai, N. Ishii, J. Itatani, M. Nakajima, H. Eisaki, A. Fujimori, and S. Shin, Phys. Rev. B {\bf 97}, 121107(R) (2018).

\bibitem{Gerber2017} S. Gerber, S.-L. Yang, D. Zhu, H. Soifer, J. A. Sobota, S. Rebec, J. J. Lee, T. Jia, B. Moritz, C. Jia, A. Gauthier, Y. Li, D. Leuenberger, Y. Zhang, L. Chaix, W. Li, H. Jang, J.-S. Lee, M. Yi, G. L. Dakovski, S. Song, J. M. Glownia, S. Nelson, K. W. Kim, Y.-D. Chuang, Z. Hussain, R. G. Moore, T. P. Devereaux, W.-S. Lee, P. S. Kirchmann, and Z.-X. Shen, Science {\bf 357}, 71 (2017).

\bibitem{Suzuki2019} T. Suzuki, T. Someya, T. Hashimoto, S. Michimae, M. Watanabe, M. Fujisawa, T. Kanai, N. Ishii, J. Itatani, S. Kasahara, Y. Matsuda, T. Shibauchi, K. Okazaki, and S. Shin, Commun. Phys. {\bf 2}, 115 (2019).

\bibitem{Mankowsky2014} R. Mankowsky, A. Subedi, M. F$\ddot{\mathrm{o}}$rst, S. O. Mariager, M. Chollet, H. T. Lemke, J. S. Robinson, J. M. Glownia, M. P. Minitti, A. Frano, M. Fechner, N. A. Spaldin, T. Loew, B. Keimer, A. Georges, and A. Cavalleri, Nature {\bf 516}, 71 (2014).

}
\end{thebibliography}

\end{document}